\documentclass[11pt]{article}
\usepackage[journal=jquant,stage=submission,logging=stdout]{jsty3-author}

\title{DC Conductance of X-shaped Majorana Interferometer reveals Non-Abelian Anyon Statistics}

\author{a,b,c}{D. Giuliano}{domenico.giuliano@fis.unical.it}{0000-0003-1642-7191}
\author*{a}{A. Nava}{a.nava@hhu.de}{0000-0002-6729-3800}
\author{a}{R. Egger}{egger@hhu.de}{0000-0001-5451-1883}
\author{d}{F. Hassler}{hassler@physik.rwth-aachen.de}{0000-0002-8903-3903}
 
\affiliation{a}{Institut f\"ur Theoretische Physik, Heinrich-Heine-Universit\"at, 40225 D\"usseldorf, Germany}
\affiliation{b}{I.N.F.N., Gruppo collegato di Cosenza, Arcavacata di Rende, I-87036  Cosenza, Italy}
\affiliation{c}{Dipartimento di Fisica, Universit\`a della Calabria Arcavacata di Rende, I-87036 Cosenza, Italy}
\affiliation{d}{Institute for Quantum Information, RWTH Aachen University, 52056 Aachen, Germany}

\abstract{We propose a four-terminal, X-shaped chiral Majorana interferometer with a central floating superconducting island, enabling the direct detection of the non-Abelian statistics of Ising anyons via the linear-response DC conductance tensor in charge transport experiments.  Here, Ising anyons are  realizable as edge vortices nucleated at Josephson line junctions defining the superconducting island, where both edge-vortex and Majorana-fermion tunneling processes can occur. We show that in such a multi-terminal Majorana interferometer, both the vacuum and the fermionic fusion channel for Ising anyons are possible. This is in contrast to two-arm interferometers, where only the vacuum fusion channel is accessible and the DC conductance contribution from edge vortices always vanishes. Using a low-energy effective theory derived via chiral bosonization, we find that in the X-shaped interferometer, the DC conductance tensor is completely isotropic, yielding a non-zero conductance when simultaneous edge-vortex and Majorana tunneling activates the fermionic fusion channel.  Apart from conductance oscillations in a gate-tunable charge parameter, which display an offset related to the anyon topological spin, measuring a finite conductance can already provide direct evidence for non-Abelian statistics in this geometry.  }

\begin{document} 

\section{Introduction}
\label{sec1}

The fractional exchange statistics of anyonic excitations has garnered a lot of attention after  recent experimental measurements of the statistical braiding phase
of Abelian anyons in fractional quantum Hall  (FQH) devices \cite{bartolomei2020,nakamura2020}.  An outstanding open challenge is to demonstrate the non-Abelian braiding statistics
of the more complex non-Abelian anyons predicted to arise, e.g., at specific FQH filling factors and in topological superconductor (SC) geometries.  Such excitations could form the basis for topological quantum information processing schemes \cite{nayak2008}.
The simplest type of non-Abelian anyon excitation is given by Ising anyons \cite{nayak2008}, which correspond to vortices harboring a zero-energy Majorana zero mode (MZM). 
When realized as excitations in the chiral one-dimensional (1D) edge modes of FQH samples or in other topologically nontrivial materials, such edge vortices (EVs) are intrinsically mobile (``flying''), and thus are attractive candidates for 
transporting quantum information \cite{beenakker2020}.  Moreover, as we discuss below, flying Ising anyons can allow experimentalists to probe the expected non-Abelian fusion rules and the associated braiding statistics of Ising anyons directly in DC charge transport experiments.
Flying Ising anyons are intimately related to their spatially localized MZM cousins \cite{barkeshli2019}, which have been discussed in recent experiments on proximitized semiconductor heterostructures and tunable quantum dot geometries and may provide practically useful Majorana qubits, see, e.g., Refs.~\cite{dvir2023,q2023,tenHaaf2024,aghaee2025,vanLoo2026,q2026,zatelli2026}.  Experimental demonstrations of Ising-anyon braiding using localized MZMs have not
yet been reported (apart from quantum simulations \cite{stenger2021,harle2023}) and are plagued by complications due to disorder effects.  Indeed, in the presence of 
disorder, conventional fermionic Andreev states may form and mimic many Majorana features \cite{prada2019}.  Flying anyons instead rely on the underlying chiral nature of the edge modes of a gapped topologically ordered bulk phase, where anyon braiding is insensitive to the presence of weak disorder \cite{bartolomei2020,nakamura2020,nayak2008,bonderson2006,bonderson2007,fendley2009,beenakker2020}. 

The conceptually simplest realization of flying Ising anyons is based on  EVs of chiral Majorana fermion edge states \cite{fendley2007,beenakker2020}. EVs can be thought of as composite objects built of a vortex in a SC (which is one-half of a fermionic flux quantum) which binds a MZM.  They constitute coreless domain walls for the chiral Majorana edge mode \cite{beenakker2020}, and are associated with a sign change at the EV position (i.e., a $\pi$ phase shift).  
In contrast to conventional SC vortices, EVs only bind a single zero-energy MZM but no additional finite-energy bound states, and thus are pure Ising anyons.
One may inject classical (deterministic) EVs into 1D chiral Majorana modes by the application of suitable voltage pulses \cite{beenakker2019b,beenakker2019a,adagideli2020,hassler2020,flor2023}. Here, we study quantum EVs
which can be dynamically nucleated, e.g., at Josephson junctions \cite{nava2024}. 
The dynamics of EVs displays non-Abelian exchange statistics accompanied by nontrivial correlation functions and multiple anyon fusion channels \cite{nilsson2010,clarke2010,hou2011,grosfeld2011,ariad2017}. For instance, 
the equal-time correlator of two EVs ($\sigma$) at long distances is given by \cite{bonderson2007,fendley2007,ariad2017}
\begin{equation}
    \langle \sigma(x)\sigma(0) \rangle\propto e^{2\pi i s_\sigma} |x|^{-2h_\sigma}, \quad s_\sigma=h_\sigma=\frac{1}{16},
\end{equation}
where the topological spin is defined as $e^{2\pi i s_\sigma}=e^{i\pi/8}$ and the conformal scaling dimension $h_\sigma$
coincides with $s_\sigma$.  The topological spin is closely tied to the non-Abelian exchange statistics of Ising anyons, and measuring its value is tantamount to detecting non-Abelian braiding \cite{kitaev2006,bonderson2007,nava2024}.

In order to probe the intriguing physics of quantum EVs, interferometric setups operated under transport conditions can offer powerful insights.  Two-arm interferometers based on chiral 1D Majorana edge modes have been studied theoretically in the absence of Josephson junctions inside the interferometer, and thus without Ising anyons. These proposed designs are based on topological insulator surfaces partially proximitized by a grounded SC, defining the interior region of the interferometer, and  by ferromagnets with different magnetization directions in the outer regions 
\cite{fu2009,akhmerov2009,nilsson2010,clarke2010,hou2011,roising2018,shapiro2021,huang2026}.
Chiral Majorana edge modes must then emerge at interfaces between gapped regions of different (SC vs. magnetic) type.
Even though clear experimental evidence for chiral Majorana edge modes is still lacking, proximity-induced superconductivity in quantum anomalous Hall insulators has been demonstrated in recent experiments, see, e.g., Refs.~\cite{amet2016,lee2017,uday2024}.
Importantly, by a proper choice of the magnetization directions of the ferromagnets, both 1D chiral Majorana edge modes will co-propagate along the same direction, in contrast to typical FQH interferometers \cite{nayak2008,beenakker2020}.  This fact leads to several important simplifications exploited in what follows.

\begin{figure}
    \centering
    \includegraphics[width=0.7\linewidth]{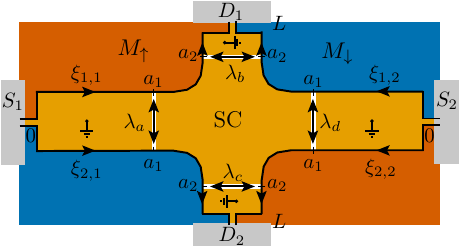} 
    \caption{Schematic setup for a symmetric X-shaped chiral Majorana interferometer formed on the surface of a topological insulator.  The surface is proximitized by a conventional superconductor (SC, orange) in the region defining the interferometer.  The outer regions are proximitized by magnetic materials of opposite magnetization  ($M_\uparrow$ and $M_\downarrow$, indicated by red and blue regions, respectively). At interfaces between magnetic materials, we have 1D chiral Dirac fermion channels (short black double lines).  At the SC-magnet interfaces, we have four neutral 1D chiral Majorana fermion modes $\xi_{n,j}(x)$ of length $L$ (black lines; arrows indicate the propagation direction), with $0<x<L$ for each mode.  These modes are populated by particle injection via 1D Dirac channels emanating from normal source electrodes $S_1$ and $S_2$. Similarly, outgoing particles are absorbed by drain electrodes $D_1$ and $D_2$.    The Dirac-Majorana conversion between Dirac and Majorana modes occurs with unit probability because of the chirality and the existence of a bulk gap. A central floating SC island with charging energy $E_C$ and offset charge parameter $n_g$ is formed by four Josephson line junctions (with Josephson energy $E_J$) at $x=a_1$ or $x=a_2$ along the Majorana modes. The outer SC regions are grounded. 
    The four Majorana fermion tunneling amplitudes $\lambda_{a,b,c,d}$ are also shown.   For  details, see Sec.~\ref{sec2a}. }
    \label{fig1}
\end{figure}

By locally connecting both arms of a linear (two-arm) interferometer \cite{fu2009,akhmerov2009} via Josephson line junctions in the SC region inside the interferometer, Majorana fermions can tunnel between different chiral Majorana edge modes. In addition, quantum EVs are nucleated at the line junctions \cite{akhmerov2009,nava2024}.
As shown in Ref.~\cite{nava2024}, by probing the low-frequency AC charge conductance of a two-arm interferometer containing a central floating SC island within the grounded SC region, one can obtain 
information about the topological spin of EVs.  However, at least for the symmetric geometry considered in Ref.~\cite{nava2024}, the DC conductance does not reveal any signature of
the braiding statistics of Ising anyons because of the simplicity of the device and the corresponding interfering trajectories.
In this paper, we show that the four-terminal geometry in figure~\ref{fig1}, representing an X-shaped Majorana interferometer, allows one to detect the non-Abelian braiding statistics of Ising anyons by direct 
measurements of electrical DC conductances in the linear-response regime.  Importantly, DC conductances are easier to measure than AC conductances. Moreover, such measurements are
 much less challenging than the shot noise and/or collision experiments proposed for detecting non-Abelian braiding statistics in the FQH regime \cite{lee2022}. In particular, conductance measurements refer to an averaged
quantity rather than to the subtle correlation functions needed, e.g., in collision experiments, see also Refs.~\cite{tang2026,dewit2026}.  For clarity, we consider a symmetric setup below.

The remainder of this article is structured as follows. In Sec.~\ref{sec2}, we describe the device in figure~\ref{fig1}
in detail, and provide a low-energy theory describing quantum transport in terms of bosonization \cite{gogolin1998}.
In Sec.~\ref{sec3}, we apply this theory to calculate the linear-response DC conductance tensor. 
We consider both  the weak-coupling  and  the strong-coupling regime. 
We find that the DC conductance tensor contains clear signatures of the non-Abelian  statistics of Ising anyons. 
In Sec.~\ref{sec4}, we summarize our results and offer an outlook.  
We put $\hbar=k_B=1$ throughout. 
 
\section{Theoretical description}\label{sec2}

In this section, we discuss the device in figure~\ref{fig1} and derive its low-energy theory 
in terms of the chiral bosonization approach \cite{gogolin1998,fendley2007,nava2024}.  
Below we assume that no localized vortices harboring MZMs are present in the 
interferometer.  However, we discuss the role of such topological defects 
in Sec.~\ref{sec4}. 

\subsection{Model}\label{sec2a}

The device in figure~\ref{fig1} features four chiral Majorana fermion modes of length $L$, 
which can be realized on a topological insulator surface which is proximitized inside the interferometer by a grounded SC with fixed phase, say, $\varphi=0$. 
A three-dimensional (3D) topological insulator, e.g., Bi$_2$Se$_3$, has a single 2D gapless Dirac fermion cone at its surface, which can be gapped out by proximitizing it with a SC or with a magnet, i.e., by depositing   either a conventional SC film or a magnetic material on the surface.   At the interfaces between gapped regions of different character, one then finds 1D chiral Dirac or Majorana edge (or interface) modes.
For simplicity, we assume that all those modes have the same edge
velocity $v$, and we use the 1D spatial coordinate $0\le x\le L$ separately for each chiral Majorana fermion mode.
Outside the interferometer region, the topological insulator surface is instead proximitized by ferromagnets with magnetization as shown in figure~\ref{fig1} such that the chirality of the Majorana  modes is as depicted.  
Assuming that the induced SC gap $\Delta$ is smaller than the 
corresponding magnetic gap, the only available low-energy modes below the gap $\Delta$ are the neutral 1D chiral Majorana modes located at the interfaces between superconducting and magnetically proximitized regions, and the 1D chiral Dirac fermions at the interfaces between different magnetic regions \cite{fu2009,akhmerov2009}.
The energy scale $\Delta$ will then also serve as bandwidth of the low-energy theory.

We next assume that away from the central floating SC island, the proximitized SC strips in figure~\ref{fig1} have a sufficiently large width, $W\gg l_c= v/\Delta$, such that the co-propagating chiral Majorana modes at opposite edges of the strip have  negligibly small overlap except at the Josephson line junctions.  Similarly, we require $L\gg l_c$.  Using typical values in proximitized topological insulator devices, say, $v\simeq 10^4$~m/s 
and $\Delta/h\simeq 100$~GHz \cite{hasan2010}, we find the typical coherence length $l_c=v/\Delta\simeq 16$~nm.  
Particles are then injected into the four Majorana edge modes via two 1D chiral Dirac channels from two normal source electrodes ($S_1$ and $S_2$) held at fixed chemical potentials $\mu_{1,2}=eV_{1,2}$. Similarly, outgoing particles are absorbed by normal drain electrodes ($D_1$ and $D_2$) held at chemical potential $\mu=0$.
One then measures the DC currents $I_{D_1}$ and $I_{D_2}$ flowing from the interferometer into the respective drain electrodes.
At each junction between a charged 1D chiral Dirac mode and a pair of neutral 1D chiral Majorana fermion modes, see figure~\ref{fig1},
a Dirac-Majorana conversion process takes place (and vice versa).  Note that charge conservation is ensured by the grounded SC condensate.  
These Dirac-Majorana conversion processes happen with unit efficiency since they are protected by chirality and by the bulk gap $\Delta$  \cite{fu2009,akhmerov2009}.
The device in figure~\ref{fig1} also contains a central floating SC island with finite charging energy $E_C$ and  dynamical (fluctuating) phase $\varphi$.  The geometric form of the island
is defined by four Josephson line junctions (at $x=a_1$ and $x=a_2$), see figure~\ref{fig1}.  
The four line junctions are engineered weak links that delimit the central
SC region in the chosen X-shaped geometry. The central region should be designed such that it has no direct superconducting short to the surrounding grounded SC film. It thus represents a floating island which is coupled to the grounded SC film through the Josephson line junctions.
We assume the same Josephson energy $E_J$ for each line junction and study the regime $E_J\gg E_C$. In this regime, fast and rare quantum phase slips, $\varphi\to \varphi\pm 2\pi$,
simultaneously affect all Josephson line junctions surrounding the central island, giving rise to the composite EV tunneling processes discussed below.
The floating central island is capacitively coupled to a gate electrode that allows one to vary the 
offset charge $n_g$ (in units of $2e$).
We note that if one gives up the geometric symmetry of the device in figure~\ref{fig1}, the theory becomes considerably more cumbersome.  We shall address the corresponding complications elsewhere.  However, the symmetric setup in figure~\ref{fig1} already provides very useful intuition on the physics of multi-terminal Majorana interferometers.

The key low-energy degrees of freedom for characterizing transport through the device in figure~\ref{fig1} are the four 1D chiral Majorana fermion modes $\xi_{n,j}(x)$ with $n,j\in \{1,2\}$. These field operators are Hermitian, $\xi^\dagger_{n,j}(x)=\xi_{n,j}^{}(x)$, and obey a fermionic anticommutation algebra,
\begin{equation}\label{MajoAlgebra}
    \{ \xi_{n,j} (x), \xi_{n',j'}(x')\} = 2\delta_{n,n'}\delta_{j,j'}\delta(x-x').
\end{equation}
We assume that all edges have the same edge velocity $v$ and the same length $L$, with $0\le x\le L$.  
In our notation, $\xi_{n,j}$ originates from the source electrode $S_j$   located at $x<-L_0$, with a Dirac-Majorana conversion at $x=0$,
where $L_0$ is the length of the 1D chiral Dirac channel in figure~\ref{fig1}.   After moving through the interferometer, the four Majorana modes are converted back to two 1D chiral Dirac channels at $x=L$, and ultimately
absorbed by the drain electrodes $D_n$ located at $x>L+L_0$.  Without Majorana and/or EV tunneling events, the low-energy Hamiltonian for chiral Majorana modes is given by
\begin{equation}\label{MajoranaHam0}
    H_0 = - \frac{iv}{2} \sum_{n,j\in\{1,2\}} \int_0^L d x \, \xi_{n,j}(x) \partial_x \xi_{n,j}(x)  .
\end{equation}
The 1D chiral Dirac fermion fields injected from the source electrodes are described by field operators $\Psi_{S_j}(x)$ with $-L_0<x<0$.  Similarly defining Dirac fields $\Psi_{D_n}(x)$ with $L<x<L+L_0$ for the outgoing channels (absorbed by the drain electrodes),    chiral Dirac fermions are modeled by 
\begin{equation}\label{HSD}
    H_{SD} = -iv\sum_{j=1,2}\int_{-L_0}^0 dx\, \Psi^\dagger_{S_j}(x)\partial_x \Psi^{}_{S_j} (x)
     -iv\sum_{n=1,2}\int_{L}^{L+L_0} dx\, \Psi^\dagger_{D_n}(x) \partial_x \Psi^{}_{D_n} (x).
\end{equation}
 Dirac-Majorana conversion processes are captured by the interface matching conditions ($n,j=1,2$) \cite{fu2009,akhmerov2009,nilsson2010}
\begin{eqnarray}\label{match1}
\xi_{1,j}(0)&=&\Psi_{S_j}^{}(0)+\Psi^\dagger_{S_j}(0),\quad \xi_{2,j}(0)= i \left( \Psi_{S_j}^{\dagger}(0)-\Psi^{}_{S_j}(0)
\right),\\  \nonumber
\xi_{n,1}(L)&=&\Psi_{D_n}^{}(L)+\Psi^\dagger_{D_n}(L),\quad \xi_{n,2}(L)= 
i \left( \Psi_{D_n}^{\dagger}(L)-\Psi^{}_{D_n}(L)\right).
\end{eqnarray}

\subsection{Chiral bosonization}

In order to study EV tunneling, it is highly advantageous to employ chiral bosonization as theoretical framework \cite{gogolin1998,fendley2007,nava2024,delft1998}. 
The four 1D chiral Majorana modes $\xi_{n,j}(x)$ with $0\le x\le L$ correspond to two neutral auxiliary
1D chiral Dirac fermion modes, which are readily bosonized in terms of two chiral
boson fields $\phi_{1}(x)$ and $\phi_2(x)$ with commutator algebra
\begin{equation}\label{bosonalgebra}
[ \phi_j(x), \phi_{j'}(x') ] = i\pi  \delta_{j,j'}\,{\rm sgn}(x-x') .
\end{equation}
For the chiral Majorana modes, we obtain the bosonized representations 
\begin{eqnarray}\nonumber
    \xi_{1,1}(x)&=& \frac{2\eta_1}{\sqrt{l_c}}  \cos\phi_1(x) ,
    \quad \xi_{2,1}(x)= \frac{2\eta_1}{\sqrt{l_c}} \sin\phi_1(x) , \\ \label{MajoBos}
    \xi_{1,2}(x)&=& \frac{2\eta_2}{\sqrt{l_c} } \sin\phi_2(x) ,
    \quad \xi_{2,2}(x)=  \frac{2\eta_2}{\sqrt{l_c}}\cos\phi_2(x) ,
\end{eqnarray}
where  $\eta_{1,2}$ are Klein factors ensuring that all anticommutation relations in Eq.~\eqref{MajoAlgebra} are satisfied and $l_c= v/\Delta$ serves as short-distance cutoff length needed in the bosonized low-energy theory \cite{delft1998}.
 The Klein factors can be chosen as Majorana fermion operators, $\eta_j^\dagger=\eta_j^{}$ and $\{\eta_{j},\eta_{j'}\}=2 \delta_{j,j'}$.
The Hamiltonian $H_0$ in Eq.~\eqref{MajoranaHam0} then takes the equivalent bosonized form 
\begin{equation}\label{H0bos}
    H_0 = \frac{v}{4\pi} \sum_{j=1,2} \int_0^L dx\,[\partial_x\phi_j(x)]^2.
\end{equation}
Similarly, the chiral Dirac fermions for $-L_0<x<0$ and $L<x<L+L_0$ are bosonized using chiral boson fields $\Phi_{S_j}(x)$ and $\Phi_{D_n}(x)$, which are defined in the respective spatial regions  and 
subject to a commutator algebra as in Eq.~\eqref{bosonalgebra}.  Explicitly, 
with Klein factors $\eta_{S_j}$ and $\eta_{D_n}$, and $l_c= v/\Delta$, one finds \cite{gogolin1998,delft1998}
\begin{equation}
    \Psi_{S_j}(x)= \frac{\eta_{S_j}}{\sqrt{l_c}}  e^{i\Phi_{S_j}(x)} ,\quad 
    \Psi_{D_n}(x)= \frac{\eta_{D_n}}{\sqrt{l_c}}  e^{i\Phi_{D_n}(x)},
\end{equation}
where $H_{SD}$ in Eq.~\eqref{HSD} takes the bosonized form
\begin{equation}\label{H0bos2}
    H_{SD}= \frac{v}{4\pi}\sum_{j}  \int_{-L_0}^0 dx\, (\partial_x\Phi_{S_j})^2 + \frac{v}{4\pi}\sum_n\int_{L}^{L+L_0} dx \,(\partial_x\Phi_{D_n})^2.
\end{equation}
Since $i\eta_1\eta_2=\pm 1$ is conserved under the full Hamiltonian considered below, we are free to choose
\begin{equation}\label{eta1eta2}
    i\eta_1\eta_2 = -1
\end{equation}
in what follows.
Finally, the matching conditions \eqref{match1} translate to ($j=1,2$)
\begin{equation}\label{match11}
    \phi_{j}(0)=\Phi_{S_{j}}(0), \quad \eta_{j}=\eta_{S_{j}}=\eta_{D_{j}},
\end{equation} 
while current conservation at $x=L$ implies  
\begin{equation}
    i \partial_x\Phi_{D_{1}}(L) = \frac{4\pi}{l_c}  \cos\phi_1(L)\: \sin\phi_2(L),
    \quad   i\partial_x\Phi_{D_{2}}(L) = \frac{4\pi}{l_c} \sin\phi_1(L) \: \cos\phi_2(L).
\end{equation} 

\subsection{Majorana fermion tunneling}

We now include Majorana fermion tunneling processes between different Majorana edges, which can occur at the four Josephson line junctions with $x=a_1$ or $x=a_2$ in figure~\ref{fig1}, where $0<a_1< a_2<L$. The assumption of negligible Majorana overlap applies to the broad superconducting strips away from the junctions. In the weak-link regions corresponding to the four junctions, the geometry and the resulting band structure intentionally bring these modes sufficiently close together such that their overlap becomes significant and produces finite local tunneling amplitudes at the engineered junctions.
Denoting these real-valued Majorana tunneling amplitudes by $\lambda_{a,b,c,d}$, see figure~\ref{fig1},
the tunneling Hamiltonian  is given by
\begin{equation}
 H_{\rm MT} = - i \lambda_a \xi_{1,1}(a_1) \xi_{2,1} (a_1) - i \lambda_b \xi_{1,2} (a_2 ) \xi_{1,1}(a_2) 
- i \lambda_c \xi_{2,1} (a_2 ) \xi_{2,2} (a_2) - i \lambda_d 
\xi_{2,2} (a_1) \xi_{1,2}(a_1) . 
\label{MTferm}
\end{equation}
Using Eq.~\eqref{MajoBos}, the bosonized representation of Eq.~\eqref{MTferm} contains a term $H_{\rm MT}^{(a,d)}$ due to $\lambda_a$ and $\lambda_d$, 
\begin{equation}\label{MajoTun1bos}
    H_{\rm MT}^{(a,d)} = \frac{\lambda_a}{\pi} \partial_x\phi_1(a_1) +\frac{\lambda_d}{\pi} \partial_x\phi_2(a_1), 
\end{equation}
which can be gauged away by a unitary transformation \cite{gogolin1998}, see below.
In addition, a term originating from $\lambda_b$ and $\lambda_c$ arises. With Eq.~\eqref{eta1eta2}, we obtain 
\begin{equation}\label{MajoTun2bos}
H_{\rm MT}^{(b,c)} =- \frac{4\lambda_b}{l_c}  \sin\phi_2(a_2)\, \cos\phi_1(a_2) + \frac{4\lambda_c}{l_c}  \sin\phi_1(a_2) \,\cos\phi_2(a_2) .
\end{equation}
We emphasize that the couplings $\lambda_{a,b,c,d}$ have dimension energy times length.

\subsection{Edge vortex tunneling}\label{sec2d}

\begin{figure}
    \centering
    \includegraphics[width=0.8\linewidth]{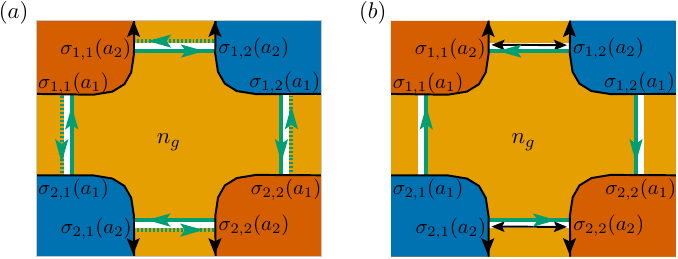} 
    \caption{Schematic illustration of composite EV tunneling processes in the X-shaped Majorana interferometer, see Secs.~\ref{sec2d} and \ref{secrg} for details.  We show the center part of figure~\ref{fig1}, with the four Josephson line junctions surrounding the floating central SC island.  Left panel: Clockwise (solid green arrowed lines) and counter-clockwise (dashed green arrowed lines) EV tunneling processes in the vacuum fusion channel.  Right panel: Example for a composite EV tunneling process in the fermionic fusion channel. Black double-arrowed lines indicate possible pairs of Majorana fermion tunneling processes.}
    \label{fig2}
\end{figure}

In addition to  Majorana fermion tunneling, let us now include EV ($\sigma)$ tunneling processes. 
Here, it is important to recall the fusion rule for Ising anyons, $\sigma\times \sigma\sim I+\psi$ \cite{nayak2008}.  While only the vacuum ($I$) fusion channel is available in a two-arm interferometer as 
studied, e.g., in Ref.~\cite{nava2024}, the multi-terminal device in figure~\ref{fig1} gives access to composite EV tunneling processes making use of the neutral fermion ($\psi$) fusion channel.
As schematically illustrated in figure~\ref{fig2},  EV tunneling processes in our geometry come in two different types, each involving a product of eight $\sigma$ operators.    First, the trivial fusion channel yields a counter-clockwise (or clockwise) tunneling process encoded by $H_{\rm EV}$.  Second,   the $\psi$ channel allows for an
additional composite EV tunneling process described by $\tilde H_{\rm EV}$.
We now specify bosonized representations for $H_{\rm EV}$ and $\tilde H_{\rm EV}$, starting with $H_{\rm EV}$.

In order to do so, we first introduce elementary EV tunneling operators for the four line junctions in figure~\ref{fig2}, 
\begin{eqnarray}\nonumber
    {\cal T}_a &=&\sigma_{1,1}(a_1) \sigma_{2,1} (a_1),\quad {\cal T}_b=\sigma_{1,2}(a_2) \sigma_{1,1} (a_2), \\ \label{Tdef} 
  {\cal T}_c&=&\sigma_{2,1}(a_2) \sigma_{2,2} (a_2),\quad  {\cal T}_d = \sigma_{2,2}(a_1) \sigma_{1,2} (a_1).
\end{eqnarray}
The indices of $\sigma(x)$ refer to the underlying  Majorana fermion modes $\xi_{n,j}(x)$.  
For instance, ${\cal T}_a$ describes EV tunneling between $\xi_{2,1}$ and $\xi_{1,1}$ at the location $x=a_1$.  Using the bosonization rules \eqref{MajoBos} with the chiral boson fields $\phi_1(x)$ and $\phi_2(x)$,  bosonized expressions for ${\cal T}_a$ and ${\cal T}_d$ have  previously been established:   With isospin operators $S^\pm=S_x\pm iS_y$, one finds ${\cal T}_a\propto \sum_\pm S^\mp e^{\pm \frac{i}{2}\phi_1(a_1)} $ and 
${\cal T}_d\propto \sum_\pm S^\pm e^{\pm \frac{i}{2}\phi_2(a_1)}$ \cite{fendley2007,nava2024}.  
We then obtain
\begin{equation}\label{HEV1}
    H_{\rm EV} \propto \Gamma  \sum_\pm  e^{\mp 2\pi i n_g} \left(\frac12\mp S_z\right)  e^{\pm \frac{i}{2}\phi_1(a_1)}  {\cal T}_b  e^{\pm \frac{i}{2}\phi_2(a_1)}  {\cal T}_c 
  \,  + \, {\rm h.c.},
\end{equation} 
with the gate-tunable offset charge parameter $n_g$. 
The conserved isospin $S_z=\pm 1/2$ in Eq.~\eqref{HEV1}
labels the total fermion parity sector, where we set $S_z=-1/2$ in what follows. 
The operators $S^\pm$ in the above relations for ${\cal T}_{a,d}$ then ensure that the vacuum fusion channel will be selected \cite{fendley2007,nava2024}.
The composite EV creation or annihilation rate $\Gamma$ is 
estimated along the lines of Refs.~\cite{schoen1990,nava2024}, 
\begin{equation}\label{Gamma}
    \Gamma \simeq \omega_p \left( \frac{E^\ast_J}{2E_C} \right)^{1/4} e^{-\sqrt{8E^\ast_J/E_C}},
\end{equation}
where $\omega_p=\sqrt{8E_J^\ast E_C}$ is the plasma frequency and $E_J^\ast=4 E_J$ is the Josephson energy of the central SC island; we assume $\omega_p\gg \Delta$. 
Quantum phase slips of the phase $\varphi$ on the central SC island occur on a time scale of order $\omega_p^{-1}$, and we treat such processes as time-local events which are effectively not affected by the dynamics of the fermionic sector. 
Note that $\omega_p$ acts as effective ultraviolet (UV) energy cutoff scale akin to $v/l_c$.  While in principle, above-gap quasiparticles could be excited by instantons for $\omega_p\gg \Delta$ \cite{flor2023}, which then could cause weak decoherence of the EV dynamics at low temperatures \cite{grosfeld2011}, the charging energy $E_C$ on the central SC island in practice suppresses such effects.  Therefore, we neglect above-gap quasiparticles in what follows.  The key energy scales (such as $\Gamma$ or temperature $T$) should be small compared to $\Delta$ for our low-energy description to be valid. 
We note in passing that instead of working with a floating central SC island, one could also choose to work with a grounded but voltage-biased SC island.  In that case, one can induce composite EV tunneling processes either at predetermined times (by applying voltage pulses \cite{beenakker2019b}) or at a constant rate (by having a constant bias voltage \cite{hassler2020}).  Under voltage-biased conditions, the expression for $\Gamma$ in Eq.~\eqref{Gamma} does not apply anymore and a broader range of rates $\Gamma$ may be accessible.

Next, we exploit that the $\sigma_{1,2}(a_2)$ operator appearing in ${\cal T}_b$, see Eq.~\eqref{Tdef}, commutes with those $\sigma$ operators contained in ${\cal T}_a$ since they belong to different chiral Majorana modes. 
A similar argument allows one to commute $\sigma_{2,1}(a_2)$, appearing in ${\cal T}_c$, with ${\cal T}_d$. 
Given the cyclic nature of the entire composite EV tunneling process, we can
thus equivalently write 
\begin{equation}
    H_{\rm EV} \propto  \Gamma e^{-2\pi i n_g}   e^{ \frac{i}{2}\phi_1(a_1) }   \sigma_{1,1}(a_2) \sigma_{2,1}(a_2)  e^{\frac{i}{2}\phi_2(a_1)}
       \sigma_{2,2}(a_2) \sigma_{1,2}(a_2) +\, {\rm h.c.},
\end{equation} 
and the remaining products of $\sigma$ operators can be bosonized as well. Indeed, with the additional spin-$1/2$ isospin operator $Y^\pm=Y_x\pm i Y_y$, where $Y_z=\pm 1/2$, we now have
$\sigma_{1,1}(a_2)\sigma_{2,1}(a_2)\propto \sum_\pm Y^\mp e^{\pm 
\frac{i}{2}\phi_1(a_2)}$ and $\sigma_{2,2}(a_2)\sigma_{1,2}(a_2)\propto \sum_\pm Y^\pm e^{\pm 
\frac{i}{2}\phi_2(a_2)}$.   Since we have chosen the sector $S_z=-1/2$ and consider the vacuum fusion channel, we here choose the conserved value $Y_z=+1/2$.
Taking into account the boson algebra \eqref{bosonalgebra}, we finally arrive at the bosonized representation
\begin{equation}\label{HEVf}
    H_{\rm EV}= \Gamma e^{-2\pi i n_g}  e^{\frac{i}{2}[\phi_1(a_1)+\phi_2(a_1)]} \: e^{-\frac{i}{2}[\phi_1(a_2)+\phi_2(a_2)]}   \, + \, {\rm h.c.}
\end{equation} 
Before discussing the term $\tilde H_{\rm EV}$,  let  us simplify notation by (i) performing a unitary transformation absorbing 
the Majorana tunneling amplitudes $\lambda_a$ and $\lambda_d$ appearing in $H_{\rm MT}^{(a,d)}$, see Eq.~\eqref{MajoTun1bos}, followed by (ii) switching to sum and difference boson fields $\phi_\pm$ instead of 
$\phi_{1,2}$. 
Step (i) is accomplished by shifting the boson fields according to \cite{gogolin1998}
\begin{equation}
\phi_1(x)\to \phi_1(x)+\frac{2\lambda_a}{v}\Theta(x-a_1),
\quad
\phi_2(x)\to \phi_2(x)+\frac{2\lambda_d}{v}\Theta(x-a_1), 
\end{equation}
with the Heaviside step function $\Theta(x)$, where we use $\Theta(0)=1/2$.
As a consequence, all effects of $\lambda_a$ and $\lambda_d$ will be accounted for by phase shifts $\varphi_\pm$ in pertinent operators, 
\begin{equation}\label{varphipm}
    \varphi_\pm = 2\frac{\lambda_a\pm \lambda_d}{v}.
\end{equation} 
From now on, we use the transformed boson fields.
Step (ii) introduces linear combinations of the boson fields,
\begin{equation}\label{bospm}
    \phi_\pm (x)= \frac{1}{\sqrt2}\left(\phi_1(x)\pm \phi_2(x) \right),
\end{equation}
which again satisfy a bosonic algebra as in Eq.~\eqref{bosonalgebra}, where 
Eq.~\eqref{H0bos} is replaced by $H_0 = \frac{v}{4\pi}\sum_\pm\int dx \, (\partial_x\phi_\pm)^2.$
With Eqs.~\eqref{eta1eta2} and \eqref{MajoTun2bos},
the remaining Majorana tunneling processes due to $\lambda_b$ and $\lambda_c$ are encoded by
\begin{equation}
    H_{\rm MT} = -2\frac{\delta \lambda}{l_c}  \sin\left[\sqrt2 \phi_+(a_2)-\varphi_+\right] 
 \,   +\, 2 \frac{\lambda_b+\lambda_c}{l_c}  \sin\left[\sqrt2 \phi_-(a_2)-\varphi_-\right] ,
\end{equation}
where we define
\begin{equation}\label{deltalambda}
\delta\lambda=\lambda_b-\lambda_c.
\end{equation}
Composite EV tunneling processes in  the vacuum fusion channel are then described by, see Eq.~\eqref{HEVf}, 
 \begin{equation}\label{HEVfinal}
    H_{\rm EV}= \Gamma \, e^{\frac{i}{2}\varphi_+ -2\pi i n_g}\, e^{\frac{i}{\sqrt2}\phi_+(a_1)}  \,  e^{-\frac{i}{\sqrt2}\phi_+(a_2)}   \,+ \,{\rm h.c.} 
\end{equation}

Next, we address composite EV tunneling processes involving the fermionic ($\psi$) fusion channel. With Eq.~\eqref{eta1eta2}, one finds  
\begin{equation}\label{tildeHEV}
    \tilde H_{\rm EV}= i\tilde \Gamma \,
    e^{-\frac{i}{2}\varphi_+ -2\pi i n_g}\,   e^{\frac{i}{\sqrt2}\phi_+(a_1)}  \,  e^{\frac{i}{\sqrt2}\phi_+(a_2)}   \,+ \,{\rm h.c.}, 
\end{equation} 
where $\tilde \Gamma$ is a pertinent rate.  We offer a detailed physical picture for 
Eq.~\eqref{tildeHEV} in Sec.~\ref{secrg}, 
where our renormalization group (RG) analysis shows that 
even for an initial value $\tilde\Gamma=0$, a finite rate $\tilde \Gamma\ne 0$ is 
dynamically generated through the interplay of 
the EV tunneling rate $\Gamma$ in Eq.~\eqref{Gamma} with the Majorana tunneling amplitude $\delta \lambda$ in Eq.~\eqref{deltalambda}.

Importantly, the full Hamiltonian separates as $H=H^{(+)}+H^{(-)}$, where $H^{(\pm)}$ only depends on the independent $\phi_\pm(x)$ boson fields and therefore $[H^{(+)},H^{(-)}]=0$.
With Eqs.~\eqref{HEVfinal} and \eqref{tildeHEV}, we find
\begin{eqnarray}\label{Hplus}
    H^{(+)}[\phi_+] &=& \frac{v}{4\pi}\int  dx \, (\partial_x\phi_+)^2 -2\frac{\delta \lambda}{l_c}  
    \sin\left[\sqrt2 \phi_+(a_2)-\varphi_+\right] \, + \, H_{\rm EV}+\, \tilde H_{\rm EV},\\
\label{Hminus}
    H^{(-)}[\phi_-] &=& \frac{v}{4\pi}\int  dx \, (\partial_x\phi_-)^2 + 2\frac{\lambda_b+\lambda_c}{l_c}  
    \sin\left[\sqrt2 \phi_-(a_2)-\varphi_-\right].
\end{eqnarray}
Note that the $H^{(-)}$ sector is blind to EV tunneling processes and can be exactly solved by refermionization techniques \cite{gogolin1998,delft1998} since it corresponds to a noninteracting fermion setting.  
The resulting calculation is equivalent to a fermionic scattering theory for chiral Majorana interferometers \cite{li2012,strubi2011}.  Since we will see in Sec.~\ref{sec3} that the $H^{(-)}$ sector does not contribute to the DC conductance tensor at all, we do not discuss this calculation here. 
 
\subsection{Renormalization group analysis}\label{secrg}

We now turn to the one-loop RG analysis of $H^{(+)}$ in Eq.~\eqref{Hplus}. 
To derive the RG scaling equations for the local couplings $\delta\lambda$, $\Gamma$, and $\tilde \Gamma$ in Eq.~\eqref{Hplus} as the UV energy bandwidth $\Lambda$ is gradually reduced from its bare (initial) value $\Lambda\simeq \Delta$ and attention is focused on the low-energy sector, we employ the standard energy-shell integration method
\cite{altland2010}.  For $\Gamma=\tilde\Gamma=\delta\lambda=0$, the $+$ boson sector has the free Euclidean action ($\tau$ denotes imaginary time) \cite{gogolin1998}
\begin{equation}
S_0 = \frac{1}{4 \pi} \int_0^{1/T} d \tau  \int_0^L d x \: 
\partial_x \phi_+ (x,\tau) \left[v \partial_x \phi_+ (x,\tau) + i \partial_\tau \phi_+ (x,\tau) \right] . 
\label{xsha.1}
\end{equation}
Introducing the chiral boson fields $w_{j=1,2}(\tau)$ at positions $x=a_1$ and $x=a_2$, and their
Fourier components $\tilde w_j(\omega_n)$ at the Matsubara frequencies $\omega_n=2\pi nT$ (integer $n$),
\begin{equation}\label{wdef}
    w_{j}(\tau)= \phi_+(a_{j}, \tau),    \quad  \tilde w_j(\omega_n)= \int_0^{1/T}d\tau \, e^{i\omega_n \tau} w_j(\tau),
\end{equation}
we next integrate over the Gaussian boson degrees of freedom away from $x=a_{1,2}$ in the partition sum for arbitrary $(\Gamma,\tilde\Gamma,\delta\lambda)$.  
In the low-temperature limit, summations over $\omega_n$ can be converted to frequency ($\omega$) integrals. With $\delta a=a_2-a_1>0$, we then obtain the effective action for the $w_j$ fields in the form 
\begin{eqnarray} \label{seffw}
&& S_{\rm eff}[w_1,w_2] =  S_{\rm eff}^{(0)}[w_1,w_2]
+ 2\Gamma
 \int d \tau  \:  \cos \left[\frac{1}{\sqrt2} \left(w_1-w_2\right) (\tau) - \frac{\pi}{4} + \frac{\varphi_+}{2} -2\pi n_g \right] \\ \nonumber 
& & - \:2 \tilde{\Gamma} 
 \int d \tau \: \sin \left[ \frac{1}{\sqrt2}\left(w_1+w_2\right) (\tau) + \frac{\pi}{4} - \frac{\varphi_+}{2} - 2\pi n_g \right]-  \frac{2\delta \lambda}{l_c} \int d \tau \: \sin\left [\sqrt{2} w_2 (\tau) - \varphi_+\right ],
\end{eqnarray}
where   integration over the bulk modes yields the action contribution
\begin{eqnarray}\nonumber
&& S_{\rm eff}^{(0)}[w_1,w_2] =
\frac{1}{4 \pi} \int_{|\omega|<\Lambda} \frac{d \omega}{2\pi}  \,
 \frac{|\omega|}{1-e^{-|\omega| \delta a/v}} \left| (\tilde w_1-\tilde w_2)(\omega) \right|^2 \\ 
 &&+ 
\frac{1}{2\pi} \int_{|\omega|<\Lambda} \frac{d\omega}{2\pi }\, |\omega| \left[ \Theta (\omega) \tilde w_1 (-\omega) \tilde w_2 (\omega) + \Theta (-\omega) \tilde w_2 (-\omega)
\tilde w_1 (\omega)\right] . \label{reno.2}
\end{eqnarray}
Note the appearance of the phase shift $\pi/4$ in the EV tunneling terms in Eq.~\eqref{seffw} which can be traced back to the bosonic algebra \eqref{bosonalgebra}. This phase shift is closely related to the topological spin of Ising anyons, see below. 

Following standard steps \cite{altland2010}, starting with the initial bandwidth $\Lambda(0)\simeq \Delta$,
the RG scheme is implemented by splitting $w_j(\tau)=w_j^{\rm sl}(\tau)+ w_j^{\rm fa}(\tau)$ into slow (sl) and fast (fa) modes, where fast modes have Fourier contributions only in a thin high-energy shell $\Lambda-d \Lambda<|\omega|<\Lambda$ of width $d\Lambda$.  
The usual RG flow parameter $\ell$ is defined by $d\ell = -d\ln \Lambda$. 
We then integrate over the fast modes and accomodate their effects on the slow-mode action by adjusting the local couplings up to one-loop order, followed by a rescaling of time and frequency. 
Performing this calculation yields the one-loop RG flow equations for $(\Gamma,\tilde \Gamma)$. We find that all Majorana tunneling amplitudes are marginal, e.g., $\frac{d}{d\ell}\delta \lambda =0$. In fact, the phases $\varphi_\pm$ in Eq.~\eqref{varphipm}, $\lambda_b+\lambda_c$ and $\delta\lambda$ stay marginal to all orders in perturbation theory and can thus be treated as constants, given by their initial values for $\Lambda(0)\simeq \Delta$.  
We now have to distinguish the cases $\Delta>v/\delta a$ vs $\Delta<v/\delta a$. 

We begin with the first case, assuming $\Delta\gg v/\delta a$.
With a phase shift from $\delta \lambda$ defined in analogy to Eq.~\eqref{varphipm}, 
\begin{equation}\label{varphic}
\varphi_{bc} = 2\frac{\delta\lambda}{v}=  2\frac{\lambda_b-\lambda_c}{v},
\end{equation}
we find the one-loop RG equations 
\begin{equation}\label{RG1}
    \frac{d\Gamma}{d\ell} =  \frac12 \left(\Gamma - \varphi_{bc} \,\tilde \Gamma\right),
    \quad
    \frac{d\tilde\Gamma}{d\ell} =\frac12\left( \tilde\Gamma - \varphi_{bc}\, \Gamma\right).
\end{equation}
We note that these RG equations for $\Delta\gg v/\delta a$ have independently been derived by using the operator product expansion (OPE) technique \cite{cardy2008} as well.  
By solving Eq.~\eqref{RG1}, we observe that the RG relevant couplings $\Gamma$ and $\tilde \Gamma$ flow toward the strong-coupling regime according to  
\begin{equation}\label{RGsolution}
   \left(\begin{array}{c} \Gamma(\ell) \\ \tilde\Gamma(\ell) \end{array}\right) 
   = e^{\ell/2} \left(\begin{array}{cc} \cosh\frac{\varphi_{bc}\ell}{2} 
   & -\sinh\frac{\varphi_{bc} \ell}{2} \\ -\sinh\frac{\varphi_{bc}\ell}{2}  &\cosh\frac{\varphi_{bc} \ell}{2}\end{array}\right) 
    \left(\begin{array}{c} \Gamma(0) \\ \tilde\Gamma(0) \end{array}\right), 
\end{equation}
with the bare (initial) values $\Gamma(0)= \Gamma$, see Eq.~\eqref{Gamma}, and $\tilde\Gamma(0)\simeq 0$.
Even for $\tilde \Gamma(0)=0$, composite EV tunneling processes in the fermionic fusion channel (described by $\tilde \Gamma$) are dynamically generated by the interplay of EV tunneling in the vacuum fusion channel (with rate $\propto\Gamma$) and Majorana fermion tunneling ($\propto \varphi_{bc}$). Here, Majorana tunneling allows for changing the fermion parity in a pair of elementary EV tunneling processes. Such a change needs to be compensated by another fermion parity change in the complementary pair in order to respect global fermion parity conservation, see figure~\ref{fig2}.  For that reason, the fermionic fusion channel is not available in a two-arm interferometer but opens up in the X-shaped geometry.
Specifically, using the OPE approach, the form of $\tilde H_{\rm EV}$ in Eq.~\eqref{tildeHEV} can be inferred from the time-ordered (${\cal T}_\tau$) operator product between $H_{\rm EV}(\tau)$ and $H_{\rm MT}^{(+)}(\tau')$ for $|\tau-\tau'|\simeq 1/\Lambda$, 
\begin{equation}\label{OPE}
{\cal T}_\tau[H_{\rm EV}(\tau) H_{\rm MT}^{(+)}(\tau')] \simeq \frac{\Gamma \delta\lambda}{v(\tau-\tau')} \frac{\tilde H_\text{EV}(\tau)}{\tilde\Gamma}
\end{equation} 
where $H_{\rm MT}^{(+)}$ is the Majorana tunneling term $\sim\delta \lambda$ in Eq.~\eqref{Hplus}.
By comparing Eq.~\eqref{OPE} with the general OPE expansion \cite{cardy2008}, 
we find that Eq.~\eqref{RG1} is indeed consistent with the emergence of $\tilde H_{\rm EV}$ in Eq.~\eqref{tildeHEV}
in the low-energy theory.
We conclude that a finite coupling $\tilde \Gamma$ will be dynamically activated by the simultaneous presence of EV tunneling ($\Gamma$) and Majorana fermion tunneling ($\delta\lambda$) as long as $\Lambda > v/\delta a$.

To establish the regime where the perturbative RG equations \eqref{RG1} are valid,
let us for simplicity take bare (initial) couplings $\delta\lambda=\tilde \Gamma=0$ for the moment, and assume equilibrium (or linear response) conditions as well as $\Gamma(0) \ll \Lambda(0)\simeq\Delta$. Then Eq.~\eqref{RGsolution} implies $\Gamma(\ell)=\Gamma e^{\ell/2}$
and $\Lambda(\ell)=\Delta e^{-\ell}$ such that at some flow parameter $\ell=\ell^\ast$ we have $\Gamma(\ell^*)=\Lambda(\ell^\ast)$ and the weak-coupling regime is left.
The corresponding temperature scale follows as $T^\ast=\Gamma(\ell^\ast)=\Lambda(\ell^\ast)=(\Gamma^2\Delta)^{1/3}$ with 
$\ell^\ast=\frac23\ln(\Delta/\Gamma)$.   The RG flow  stops once  $\Lambda(\ell)$ reaches ${\rm max}(T, v/\delta a)$,
and the weak-coupling regime described by Eqs.~\eqref{RG1} and \eqref{RGsolution} is realized for
\begin{equation}\label{crossover1}
    \Delta \gg \tilde\varepsilon > T^* = (\Gamma^2\Delta)^{1/3},\quad
    \tilde\varepsilon={\rm max}(T,v/\delta a).
\end{equation}
This condition is satisfied in two important limits.  First, for $\delta a>v/T$, it holds at high temperatures in the regime  $T^\ast<T\ll \Delta$.  Second, for $v/\Delta\ll \delta a<v/T$, the RG flow stops at $\tilde \ell  \simeq \ln(\Delta \delta a/v)$.   

Let us now allow for an initial value $\delta \lambda\ne 0$, 
and estimate the final value $\tilde \Gamma_\text{eff}$ in the weak-coupling regime.
To that end, we assume that $\varphi_{bc}\ell$ is small, i.e., 
$\ln(\Delta/\tilde \varepsilon)\,\delta \lambda \ll v$, and
evaluate the RG flow up to the scale $\tilde\ell = \ln(\Delta/\tilde\varepsilon)$. 
This yields the effective coupling
\begin{equation}\label{tildeGammaestimate}
  \tilde \Gamma_\text{eff} = \tilde \Gamma(\tilde \ell) \simeq -\frac{\Gamma \delta \lambda}{v} \left(\frac{\Delta}{\tilde \varepsilon}\right)^{1/2} \ln(\Delta/\tilde \varepsilon),
\end{equation} 
which serves as estimate for $\tilde \Gamma$ in our discussion below.

Let us now briefly address the complementary regime $\delta a < v/\Delta$, where the OPE method is not reliable anymore and only the energy-shell approach yields correct one-loop RG equations. We find
\begin{equation}\label{RG2}
    \frac{d\Gamma}{d\ell} =  \Gamma, \quad  \frac{d\tilde\Gamma}{d\ell} =  - \varphi_{bc}\, \Gamma.
\end{equation}
Importantly, $\Gamma$ now becomes more strongly relevant, $\Gamma(\ell)=e^{\ell}\Gamma(0)$, while $\tilde \Gamma$ is only marginally relevant because of the mixing between $\Gamma$ and $\delta\lambda$. 
Physically, for $\delta a\to 0$, we find that $\tilde \Gamma$ simply reduces to Majorana fermion tunneling processes since EV tunneling processes encoded by $\Gamma$ fuse to the identity.  
For a nontrivial Ising anyon fusion in the fermionic channel to happen, EV excitations therefore ``need space'' in the  sense that $\delta a> v/\Delta$  is required.
In the remainder of this work, we therefore assume the latter condition to hold.

Outside the regime \eqref{crossover1}, one enters a strong-coupling regime where the 
couplings $\Gamma$ and $\tilde\Gamma$ flow to strong coupling.  This regime is not described by Eq.~\eqref{RG1} anymore.  We discuss the conductance tensor in the latter regime in Sec.~\ref{sec3b}.

\section{DC conductance tensor}
\label{sec3}

In this section, we address charge transport in the X-shaped Majorana interferometer, see figure~\ref{fig1}, in the linear response regime of small applied voltages, $|eV_{1,2}|\ll T$.  
Let us first define operators $I_{S_j}$ (with $j\in \{1,2\}$) describing the charge currents injected into the interferometer from the respective source electrode via 1D Dirac channels,
\begin{equation} \label{ISj1}
    I_{S_j} = ev  \, \Psi^\dagger_{S_j}(0) \Psi^{}_{S_j}(0)   = - i \frac{ev}{2\pi} \partial_x\phi_j(0).
\end{equation}
In the second step, we used bosonization and the matching condition \eqref{match11}. Switching to $\phi_\pm(x)$ in Eq.~\eqref{bospm}, the operators \eqref{ISj1} decompose accordingly, 
\begin{equation}\label{ISj}
    I_{S_1}=I^{(+)}_{S} + I^{(-)}_S, \quad  I_{S_2}= I^{(+)}_{S} - I^{(-)}_S,\quad 
    I_S^{(\pm)} = -i\frac{ev}{2\pi\sqrt2} \partial_x\phi_\pm(0)\,.
\end{equation}
Similarly, using Eqs.~\eqref{eta1eta2} and \eqref{MajoBos}, the operators ($n\in \{1,2\}$)
\begin{equation} \label{IDn1}
 I_{D_n} = ev  \, \Psi^\dagger_{D_n}(L) \Psi^{}_{D_n}(L)  
\end{equation}
describe outgoing charge currents absorbed by the respective drain electrode. 
Bosonizing those operators, see Eqs.~\eqref{MajoBos}, \eqref{eta1eta2}, and \eqref{bospm},
we find that they again decompose in the $\pm$ boson sectors,
\begin{equation}\label{IDn}
    I_{D_1} = I^{(+)}_{D} - I^{(-)}_D, \quad 
    I_{D_2}= I^{(+)}_{D} + I^{(-)}_D,\quad 
    I_D^{(\pm)}= \frac{ev}{l_c}  \sin\left[\sqrt2 \phi_\pm(L) - \varphi_\pm\right]  .
\end{equation}
Just like the Hamiltonian, see Eqs.~\eqref{Hplus} and \eqref{Hminus}, we conclude that 
also the charge currents can be fully separated into $\pm$ contributions.
As discussed below, the $-$ sector, which can be treated exactly by refermionization techniques \cite{gogolin1998}, does not contribute to
steady-state expectation values of the DC current,  $\left\langle I_{S}^{(-)}\right\rangle=\left\langle I_D^{(-)}\right\rangle =0$. 
  Hence low-energy DC transport through the interferometer in figure~\ref{fig1} is solely determined by the $+$ sector.  

Here, we study the DC conductance tensor ${\bf G}$ with matrix elements $G_{jj'}$, which for the device in figure~\ref{fig1} is a $2\times 2$ matrix connecting the outgoing currents $\left\langle I_{D_j} \right\rangle$, see
Eq.~\eqref{IDn}, to the applied voltages $V_1$ and $V_2$,
\begin{equation}\label{condtens}
    \left( \begin{array}{c} \langle I_{D_1} \rangle \\ \langle I_{D_2} \rangle \end{array} \right)
    = \left( \begin{array}{cc} G_{11} & G_{12} \\ G_{21} & G_{22} \end{array} \right)
    \left( \begin{array}{c} V_1  \\ V_2 \end{array} \right).
\end{equation}
In the absence of EV tunneling, i.e., for $\Gamma=\tilde\Gamma=0$, we find  
$\left\langle I_{D_{1,2}} \right\rangle = 0$ for arbitrary values of $V_1$ and $V_2$, and therefore even the nonlinear
DC conductance tensor vanishes.  One can rationalize this result by noting that 
every electron or hole injected from $S_1$ or $S_2$ splits into a pair of Majorana fermion modes.  As a result of the uncorrelated nature of fermions originating from $S_1$ and $S_2$, there cannot be 
any charge transport since no path exists where two Majorana fermions originate from the same source and recombine 
in the same drain. One therefore finds $\left\langle I_{D_{1,2}}\right\rangle=0$ for $\Gamma=\tilde\Gamma=0$
in a symmetric X-interferometer, and hence a vanishing DC conductance tensor.  This feature is known to be a defining characteristic for the Majorana nature of these chiral edge states \cite{strubi2011,li2012}.
To obtain it on a formal level, we first note that chiral Majorana modes are only defined up to a $\mathbb{Z}_2$ gauge factor. For instance, from Eq.~\eqref{match1}, we have at $x=0$ the conditions 
$2 \Psi_{S_j}^\dagger(0) \Psi_{S_j}^{}(0)  - 1  = i \xi_{1,j}(0) \xi_{2,j}(0)$.  We can now apply a $\mathbb{Z}_2$ gauge factor $c_j=\pm 1$ (for each source) to the Majorana modes, $\xi_{1,j}(x)\to c_j \xi_{1,j}(x)$ and
$\xi_{2,j}(x)\to c_j \xi_{2,j}(x)$, without changing the physics.  Without Majorana tunneling processes, the current  $I_{D_{n}}$ entering a given drain electrode $D_n$ must clearly come from two Majorana modes that 
originate from different source electrodes, see figure~\ref{fig1}.  As a consequence, $I_{D_n}$ will be proportional to $c_1c_2$ under $\mathbb{Z}_2$ gauge transformations, and gauge invariance forces the steady-state DC 
currents $\langle I_{D_n} \rangle$ to vanish \cite{li2012}. For our symmetric geometry, since Majorana tunneling processes can be included through a unitary basis rotation, this argument remains valid even for finite amplitudes $\lambda_{a,b,c,d}$.

We focus on the linear-response regime, where $H'=-\sum_{j} eV_j N_{S_j}$ is treated in lowest-order perturbation theory.  Here, $N_{S_j}$ refers to the particle number in the respective source electrode, with $e\dot N_{S_j}=-I_{S_j}$. 
Using standard arguments \cite{altland2010}, we obtain a Kubo formula for the DC conductance matrix elements,
\begin{equation}\label{kubo1}
    G_{jj'} = - \lim_{\omega_n\to 0} \frac{K_{jj'}(\omega_n)}{\omega_n},
\end{equation}
where the analytic continuation $i\omega_n\to \omega+i0^+$ followed by $\omega\to 0$ is understood.
The kernel $K_{jj'}(\omega_n)$ in Eq.~\eqref{kubo1} follows from the time-ordered imaginary-time correlation function
of the drain current $I_{D_j}(\tau)$ and the source current $I_{S_{j'}}(0)$,
\begin{equation}\label{Kjjp}
    K_{jj'}(\omega_n)= \int_0^{1/T} d\tau\, e^{i\omega_n \tau} 
    \left\langle {\cal T}_\tau \, I_{D_j}(\tau) I_{S_{j'}}(0)  \right\rangle.
\end{equation}
The above arguments for $\Gamma=\tilde\Gamma=0$ and the absence of EV tunneling processes in $H^{(-)}$, see Eq.~\eqref{Hminus}, also imply that the $\phi_-$ sector does not contribute to 
the corresponding current-current correlation function,
\begin{equation}\label{curminus}
    \left\langle {\cal T}_\tau\,I_D^{(-)}(\tau) I_S^{(-)}(0) \right\rangle =0.
\end{equation}
Since $H^{(+)}$ and $H^{(-)}$ are decoupled, finite contributions then only arise from the correlation function 
$\left\langle {\cal T}_\tau\,I^{(+)}_{D}(\tau) I^{(+)}_{S}(0)\right\rangle.$  This conclusion holds for arbitrary $\Gamma$ and $\tilde \Gamma$, and implies that the linear-response
DC conductance tensor is expressed in terms of a single conductance $\tilde G$ for the entire
low-energy regime (including both the weak-coupling and the strong-coupling limits),
\begin{equation}\label{Gtensor}
{\bf G} = \tilde G \left( \begin{array}{cc} 1 & 1\\ 1& 1 \end{array} \right).
\end{equation}
The conductance $\tilde G$ is here given by
\begin{equation}\label{Kubo}
    \tilde G= - \lim_{\omega_n\to 0} \frac{K(\omega_n)}{\omega_n},\quad
    K(\omega_n) = \int_0^{1/T}d\tau\, e^{i\omega_n\tau} 
    \left\langle {\cal T}_\tau \, I^{(+)}_{D}(\tau) I^{(+)}_{S}(0) \right\rangle.
\end{equation}
We emphasize that for $\Gamma=\tilde \Gamma=0$, the conductance vanishes, $\tilde G=0$, irrespective of the values of $\lambda_{a,b,c,d}$.  Measuring a finite DC conductance under low-energy conditions (where our theory applies) thus directly serves to establish signatures of
quantum EV tunneling processes.

We now analyze Eq.~\eqref{Kubo} separately in the weak-coupling regime and in the strong-coupling regime as defined in Sec.~\ref{secrg}.

\subsection{Weak-coupling regime}\label{sec3a}

In the weak-coupling regime as defined by Eq.~\eqref{crossover1}, a perturbative treatment of the local tunneling processes is sufficient in order to evaluate Eq.~\eqref{Kubo} \cite{delft1998,campagnano2016}. To lowest nontrivial order,
using the bosonized expressions for $I_S^{(+)}$ and $I_D^{(+)}$ in Eqs.~\eqref{ISj} and \eqref{IDn}, respectively, we 
find that only the term $\propto \tilde\Gamma$ gives a finite contribution,
\begin{equation}\label{kubopert}
K(\omega_n)\simeq - 2\pi e^2 \tilde \Gamma \sqrt{\frac{\pi T }{\Delta}} e^{\omega_n L/v}  
 \frac{\sinh(\omega_n\delta a/v)}{\sinh^{1/2}(\pi T \delta a/v)} \sin\left( \frac{\pi}{4}-\frac{\varphi_+}{2}+2\pi n_g\right)  ,
\end{equation}
where we have  used $l_c=v/\Delta$.  We emphasize that EV tunneling in the vacuum fusion channel alone is not able to produce a finite DC conductance, similar to what happens in the two-arm interferometer studied in Ref.~\cite{nava2024}.  The fermionic fusion channel for Ising anyons, opening up in our multi-terminal geometry for sufficiently large $\delta a$, is of critical importance in establishing a finite conductance.

Analytic continuation to real frequencies and sending $\omega\to 0$ then gives the DC conductance tensor in Eq.~\eqref{Gtensor}. With the conductance quantum $G_0=e^2/(2\pi \hbar)$ and using the estimate in Eq.~\eqref{tildeGammaestimate} for $\tilde \Gamma$, we obtain 
  \begin{equation}\label{tildeG}
    \frac{\tilde G}{G_0} \simeq - 4\pi^2 \frac{\Gamma \delta \lambda}{v}  \left(\frac{\delta a}{v\tilde \varepsilon}\right)^{1/2} \ln\left(\frac{\Delta}{\tilde \varepsilon}\right)
  \left(\frac{\pi T \delta a /v}{\sinh(\pi T \delta a/v)}\right)^{1/2}  
  \sin\left( \frac{\pi}{4}-\frac{\varphi_+}{2}+2\pi n_g\right),
\end{equation}  
where $\tilde \varepsilon={\rm max}(T,v/\delta a)$. 
We emphasize that Eq.~\eqref{tildeG} holds under the conditions in Eq.~\eqref{crossover1}.

At this point, several observations are in order.  First, the phase shift $\pi/4$ in Eq.~\eqref{tildeG} is a manifestation of the topological spin.  In fact, it is twice the braiding phase $\pi/8$ discussed in Sec.~\ref{sec1} 
because $\tilde H_{\rm EV}$ is composed by two pairs of elementary EV tunneling events, see figure~\ref{fig2}.  The offset charge $n_g$ can be experimentally varied by changing a gate voltage and its effects can be measured
by Coulomb blockade spectroscopy.  Similarly, the phase $\varphi_+$, see Eq.~\eqref{varphipm}, arises from Majorana tunneling amplitudes which can be tuned by modulating the Josephson energy of the corresponding line junctions 
by finger gates.  By careful calibration protocols, one can then extract the phase shift $\pi/4$ in Eq.~\eqref{tildeG} which provides a signature for non-Abelian braiding statistics of Ising anyons in the DC conductance tensor.

Another hallmark for Ising anyon statistics follows from the fact that $\tilde G=0$ unless we have both a finite composite EV tunneling rate $\Gamma$ and a finite Majorana tunneling amplitude $\delta \lambda$.   Only then 
$\tilde \Gamma\ne 0$ is possible, see Eq.~\eqref{OPE}, and a finite value of the DC conductance tensor can be expected.  This fact establishes a direct connection to non-Abelian statistics since the latter can be understood in terms of having multiple fusion channels for anyons. In particular, figure~\ref{fig2} and our calculations illustrate that for small $\delta a$, only the vacuum fusion channel matters for EVs, where  transport currents are expected to vanish. In particular, for $\delta a\to 0$, we find $\tilde G\propto \sqrt{\delta a}\to 0$ since then $\tilde H_{\rm EV}$ effectively describes only Majorana fermion tunneling, which cannot result in a finite conductance as explained above. For larger $\delta a$, however, the fermionic fusion channel opens up, as characterized by $\tilde \Gamma\ne 0$, and a finite DC conductance emerges.  

\begin{figure}
    \centering
    \includegraphics[width=\linewidth]{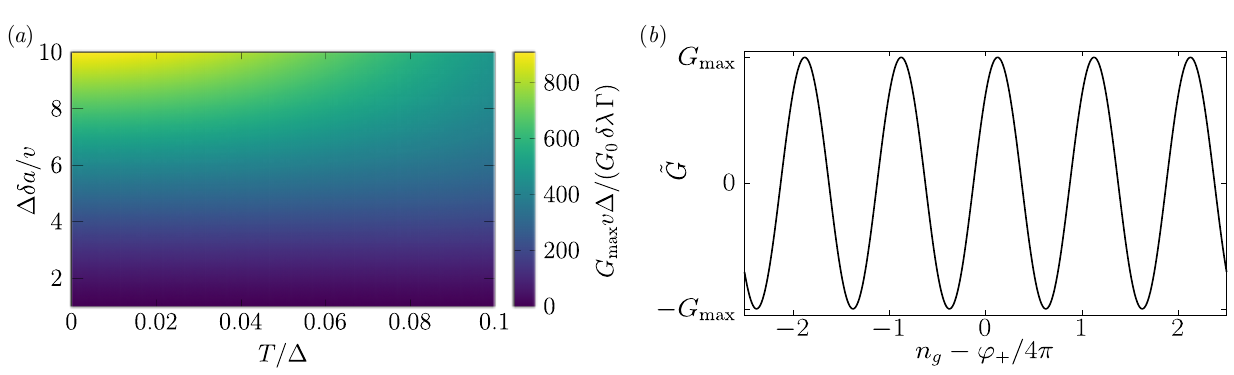} 
    \caption{Linear-response DC conductance $\tilde G = G_\text{max} \sin(\pi/4 -\varphi_+/2 + 2\pi n_g)$  in Eq.~\eqref{tildeG}, which determines the conductance tensor \eqref{Gtensor} in
    the weak-coupling regime of an X-shaped Majorana interferometer, see Eq.~\eqref{crossover1}. ($a$) 2D color-scale plot for the conductance amplitude $G_\text{max}$ in the plane spanned by temperature $T$ (in units of $\Delta$) and $\delta a=a_2-a_1$ (in units of $l_c=v/\Delta)$, see figure~\ref{fig1}.  As $T$ decreases and $\delta a$ increases, the growth of $G_\text{max}$ indicates a crossover toward the strong-coupling regime.  ($b$) The conductance $\tilde G$ oscillates as a function of the effective gate charge $n_g - \varphi_+/4\pi$ with an offset $1/8$ due to the topological spin of Ising anyons. }
    \label{fig3}
\end{figure}

Third, Eq.~\eqref{tildeG} predicts a specific dependence of $\tilde G$ on the geometry (encoded by the dependence on $\delta a$) and on temperature $T$.   
In particular, for $T\gg v/\delta a$ (but $T\ll \Delta$), the conductance becomes exponentially small due to thermal suppression effects.  For small $\tilde\varepsilon={\max}(T,v/\delta a)$, however
Eq.~\eqref{tildeG} predicts (up to logarithmic corrections) a $\tilde G\propto \tilde\varepsilon^{-1/2}$
dependence, which is due to the fact that $\tilde H_{\rm EV}$ is a relevant perturbation.  
One therefore expects that at low temperatures and for large $\delta a$, the conductance $\tilde G$ increases and eventually enters the strong-coupling regime.

We illustrate the above properties of the conductance $\tilde G$ in Eq.~\eqref{tildeG} 
in figure~\ref{fig3}, where we show both the dependence of the maximal conductance
on $T$ and $\delta a$ as a color-scale plot, see panel $(a)$, and the conductance oscillations as a function of the gate-tunable parameter $n_g-\varphi_+/4\pi$, with an offset $1/8$ due to the topological spin of Ising anyons, see panel $(b)$.  Last but not least, the conductance tensor \eqref{Gtensor} is completely isotropic, which provides yet another nontrivial experimental check of our theory.

\subsection{Strong-coupling regime}\label{sec3b}

Next, we explore the strong-coupling regime realized once the RG flow leaves the regime \eqref{crossover1}.
This happens in particular at very low temperatures. For concreteness, we assume $\Delta\gg v/\delta a$, $\Gamma(0)=\Gamma$, $\tilde \Gamma(0)=0$, and $\varphi_{bc}>0$, where Eq.~\eqref{RGsolution} predicts a flow toward a strong-coupling fixed point with $\Gamma\to \infty$ and $\tilde \Gamma\to -\infty$. Even though this limit is not 
described by the perturbative RG approach, the effective action $S_{\rm eff}[w_1,w_2]$ in Eq.~\eqref{seffw} can nonetheless be used to expand around this putative fixed point in order to study its stability and the associated DC conductance tensor.  

In a first step, we neglect the marginal Majorana tunneling processes $\propto \delta \lambda$ in Eq.~\eqref{seffw}. In addition, we also neglect the effect of the applied voltages $V_{1,2}$.  
For $\Gamma\to \infty$ and $\tilde \Gamma\to -\infty$,  the very strong pinning potentials exerted by $H_{\rm EV}$ and $\tilde H_{\rm EV}$ imply that only static solutions are possible, $w_{1,2}(\tau)=\bar w_{1,2}$, see also
Eq.~\eqref{wdef}.  
These minimizers of the pinning potentials must satisfy the conditions (with $n,m\in \mathbb{Z}$)
\begin{eqnarray}\nonumber
    \frac{\bar w_1-\bar w_2}{\sqrt2} -\frac{\pi}{4} +\frac{\varphi_+}{2}-2\pi n_g &=& 2\pi n+ \pi + \rho_- ,\\  \label{pinning}
    \frac{\bar w_1+\bar w_2}{\sqrt2} +\frac{\pi}{4} -\frac{\varphi_+}{2}-2\pi n_g &=& 2\pi m+ \frac{\pi}{2} + \rho_+ ,
\end{eqnarray}
where for now, $\rho_-=\rho_+=0$.
At large but finite values of $\Gamma$ and $|\tilde \Gamma|$,  we also need to include the effects of the Majorana tunneling term $\propto \delta \lambda$. Moreover,
to obtain information about charge transport, the applied voltages $V_{1,2}$ have to be incorporated. Since we are interested in the DC conductance tensor, the drain currents can effectively be evaluated at $x=a_2$ such that  
$I_D^{(+)}(\tau)= \frac{ev}{l_c} \sin\left[\sqrt2 w_2(\tau)-\varphi_+\right]$, see Eq.~\eqref{IDn}.

In order to compute $\tilde G$ in Eq.~\eqref{Gtensor}, it suffices to apply a uniform voltage bias $V$. The coupling to the voltage sources is
then given by $H'=-V\Delta Q^{(+)}$, with the charge imbalance $\Delta Q^{(+)}=\int_{-\infty}^{a_1} dx \rho(x)= ew_1/(2\sqrt2\, \pi)$.  Considering again only static solutions, the steady-state current at finite $V$ follows as
\begin{equation}\label{currentstrong}
    \left\langle I_{D}^{(+)} \right\rangle_V =- \frac{ev }{2T} \frac{\partial \ln {\cal Z}}{\partial \delta \lambda},
\end{equation} 
where the partition function ${\cal Z}$ is obtained by summing only over static solutions $w(\tau)=\bar w$ but in the presence of $H'$.  This implies a summation over all  $(n,m)$, where ${\cal Z}$ yields a current-carrying steady-state solution for $V\ne 0$.
The relation \eqref{currentstrong} uses the fact that $\delta\lambda$ is essentially the source field of $I_D^{(+)}$, see Eq.~\eqref{seffw}.
To account for deviations from the pinning conditions \eqref{pinning} because of $\delta \lambda$ and/or $V$, 
we allow for small but finite values of $\rho_\pm$ in Eq.~\eqref{pinning}.  Inserting Eq.~\eqref{pinning} into the effective action $S_{\rm eff}[\bar w_1,\bar w_2]$ in \eqref{seffw}, where Eq.~\eqref{reno.2} 
simplifies to $S_{\rm eff}^{(0)}=\frac{v}{4\pi T\delta a} (\bar w_1-\bar w_2)^2$, we find that a given choice for $\rho_\pm$ at fixed $(n,m)$ is characterized by the effective potential 
\begin{equation}
    V_{n,m}(\rho_+,\rho_-) = V^{(+)}_n(\rho_+)+ V^{(-)}_{n,m} (\rho_-).
\end{equation}
The respective potentials are given by (with $\tilde \Gamma\simeq -\Gamma$)
 \begin{eqnarray}
    V_{n}^{(+)} (\rho_+) &=& \Gamma  \rho_+^2 - \left( \frac{2\delta \lambda}{l_c} + \frac{eV}{4\pi}\right) \rho_+ + \frac{v}{2 \pi \delta a } \left[ 2 \pi \left(
n + n_g + \frac12\right) + \frac{\pi}{4} - \frac{\varphi_+}{2}\right]^2, \nonumber \\
V_{n,m}^{(-)} (\rho_- ) &=& \left(  \Gamma   + \frac{v}{2 \pi \delta a} \right)\rho_-^2 - 
\left\{-\frac{2 \delta \lambda}{l_c} + \frac{eV}{4\pi} + \frac{v}{2 \pi \delta a}\left[ 2 \pi \left(n + n_g + \frac12 \right) 
+ \frac{\pi}{4}- \frac{\varphi_+}{2}\right] \right\}\rho_- \nonumber \\
&-& \frac{eV}{2}\left(n+m + 2 n_g  + \frac{3}{4}\right ). 
\end{eqnarray}
The drain current \eqref{currentstrong} for given $\rho_\pm$ at voltage $V$ can then be written as
 \begin{equation}\label{currentstrong2}
\left\langle I_{D}^{(+)} \right\rangle = \frac{ev}{l_c}  \left[ \left.\sin(\rho_--\rho_+)\right|_{V} -  \left.\sin(\rho_--\rho_+)\right|_{V=0} \right]. 
\end{equation} 
For given $n$ and $m$, the minimizers are given by 
\begin{equation}
    \rho_{+;n} = \frac{\delta \lambda}{2l_c\Gamma} + \frac{eV}{8\pi\Gamma},\quad 
\rho_{-;n,m} = \frac{-\frac{\delta \lambda}{l_c} + \frac{eV}{8\pi} +\frac{v}{2\pi \delta a} \left[
2 \pi \left(n + n_g  + \frac{1}{2}\right) + \frac{\pi}{4}- \frac{\varphi_+}{2}\right]}{\Gamma + \frac{v}{2 \pi \delta a}},
\end{equation} 
and the potential $V_{n,m}^{\rm min}=V_{n,m }(\rho_{+;n},\rho_{-;n,m})$ evaluated at the minimizers is 
\begin{eqnarray}
    V_{n,m}^{\rm min} &=& - \frac{ \left(\frac{2 \delta \lambda}{l_c} + \frac{eV}{4\pi} \right)^2}{2\Gamma} + \frac{v}{2 \pi \delta a} 
\left[2 \pi \left(n + n_g +\frac{1}{2} \right) + \frac{\pi}{4} - \frac{\varphi_+}{2}\right]^2 \\ \nonumber 
&-& \frac{\left\{- \frac{2 \delta \lambda}{l_c} + \frac{eV}{4\pi} + \frac{v}{\pi \delta a} \left[ 2 \pi \left(n + n_g  + \frac{1}{2} \right) + \frac{\pi}{4}
- \frac{\varphi_+}{2} \right]\right\}^2}{2 \left( \Gamma + \frac{v}{2 \pi \delta a} \right)}  
- \frac{eV}{4 \pi} (n+m + 2 n_g ) - \frac{3 eV}{8}.
\end{eqnarray}
Clearly, the $V_{n,m}^{\rm min}$ are unbounded with respect to the integer variable $m$.  However, the current in Eq.~\eqref{currentstrong2} does not depend on $m$ at all, and we can thus take 
$e^{- V_{n,m=0}^{\rm min}/T}$ as Boltzmann weight when computing its value.  

In the large-$\Gamma$ limit, we then find 
\begin{equation}
    \left\langle I_{D}^{(+)} \right\rangle \simeq \frac{v}{4l_c\Gamma} \left(1- \frac{1}{1+\frac{v}{2\pi \Gamma\delta a}}  \right) G_0 V.  
\end{equation}
Using $l_c=v/\Delta$, we obtain the linear-response DC conductance tensor \eqref{Gtensor} in the strong-coupling limit with $\tilde G$ given by
\begin{equation}\label{dcstrongcoupl}
    \frac{\tilde G}{G_0} \simeq \frac{\Delta}{4\Gamma} \frac{1}{1+\frac{2\pi \Gamma\delta a}{v}}.
\end{equation}
In the deep strong-coupling limit, $\tilde G$ becomes insensitive to changes in $n_g$ or $\varphi_+$, and thus the topological spin cannot be detected from the DC conductance anymore. Even though $\delta \lambda$ does not show up explicitly in Eq.~\eqref{dcstrongcoupl} anymore, we note that our assumption $\tilde \Gamma=-\Gamma$ implicitly asserts that the fermionic fusion channel remains active.  The finite result for $\tilde G$ in Eq.~\eqref{dcstrongcoupl} can therefore also provide evidence for the non-Abelian statistics of Ising anyons. 

On top of the static field configurations, for large but finite values of $\Gamma$ and $|\tilde\Gamma|$, time-dependent instanton and anti-instanton field configurations interpolating between different potential minima should be taken into account \cite{altland2010}.  To study the role of instantons, let us, e.g., consider an instanton--anti-instanton pair in the field $w_2(\tau)$, where $(n,m)\to (n\mp 1,m\pm 1)$. Neglecting the finite width of the (anti-)instanton, we have 
\begin{equation}\label{w2ins}
    w_2(\tau)= \bar w_2 + 2\sqrt2 \, \pi \left[\Theta(\tau-\tau_1)-\Theta(\tau-\tau_2)\right] ,
\end{equation}
with $\tau_1$ and $\tau_2$ referring to the (anti-)instanton positions in imaginary time.  Inserting Eq.~\eqref{w2ins} and $w_1(\tau)=\bar w_1$ into the action \eqref{seffw}, we find that instanton--anti-instanton 
separation time $|\tau_1-\tau_2|$ affects the weight $e^{-S_{\rm eff}^{(+)}}$ in the partition function as
\begin{equation}
    e^{-S_{\rm eff}^{(+)}} \propto |\tau_1-\tau_2|^{-4}.
\end{equation}
The scaling dimension of a single instanton--anti-instanton creation operator is therefore $d=2$, and such processes are therefore strongly irrelevant near the strong-coupling fixed point.  Similar conclusions are reached when
considering instanton--anti-instanton configurations in $w_1(\tau)$.  We conclude that time-dependent instanton corrections to the DC conductance \eqref{dcstrongcoupl} in the strong-coupling regime are highly irrelevant, and will therefore be very small and unlikely to be detectable in experiments.  For that reason, even though one can compute the corresponding conductance contributions by following the steps outlined in Ref.~\cite{nava2024}, we refrain from the corresponding calculations in this paper.

\section{Discussion and Conclusions}\label{sec4}

In this paper, we have introduced and analyzed an X-shaped Majorana interferometer geometry where four 1D chiral Majorana modes connect two source and two drain electrodes.  In the presence of a central floating SC island, different chiral Majorana modes are connected by 
Majorana fermion tunneling processes as well as by composite EV tunneling events.  Importantly, EVs realize Ising anyons which co-propagate with the chiral Majorana fermions and are subject to non-Abelian braiding statistics.  While in the simpler two-arm geometry \cite{fu2009,akhmerov2009,beenakker2020,nava2024}, Ising anyons can only fuse in the vacuum fusion channel, the multi-terminal interferometer introduced here also allows for anyon fusion processes in the fermionic channel.  
This important and distinct feature has far-reaching consequences.  In particular, in the absence of either EV tunneling or Majorana tunneling events, and at low energy scales compared to the induced pairing gap $\Delta$, steady-state currents flowing into the drain electrodes must vanish altogether.  Only in the simultaneous presence of Majorana fermion and EV tunneling processes, the fermionic fusion channel will be activated and can generate a finite DC conductance.  A remarkable consequence of our theory is that the mere observation of a finite low-energy conductance already represents a signature for the non-Abelian statistics of Ising anyons, which is also manifest in a $\pi/4$ phase shift of the conductance oscillations with respect to a gate-tunable charge parameter. This phase shift is connected to the topological spin of Ising anyons.  Moreover, our X-shaped Majorana interferometer features a completely isotropic conductance tensor, see Eq.~\eqref{Gtensor}, which is parametrized by a single conductance $\tilde G$ at low energies.  It is a very interesting challenge to experimentally test these predictions.  

Above we have assumed that no localized topological defects (e.g., vortices harboring MZMs) are present inside the interferometer.  For a two-arm interferometer, such defects can affect the conductance, in particular the sign thereof, see, e.g., Refs.~\cite{fu2009,akhmerov2009,nilsson2010,nava2024}.  In our X-shaped interferometer, similar effects can occur, e.g., for two bulk defects inside the central SC island, where the fermionic state built from these two MZMs can either be empty or occupied. In the latter case, we have a shift of the offset charge $n_g\to n_g+ 1/2$ (in units of $2e$) because this corresponds to having an unbound electron on the floating SC island.  In the weak-coupling regime, this shift implies a sign change of the DC conductance tensor, see Eqs.~\eqref{Gtensor} and \eqref{tildeG}. 
However, the same effect can also be caused by topologically trivial localized subgap Andreev states on the central island.  If such a state becomes filled, we also have $n_g\to n_g+1/2$, and thus 
a sign change of the conductance.

In this paper, we have assumed a symmetric interferometer geometry.
Geometric asymmetries introduce unequal propagation times and modify interference
phases and amplitudes, so that the particularly simple isotropic conductance tensor derived
here need not persist. However, they do not by themselves remove the vacuum and fermionic
fusion channels available in the device. We therefore expect the finite-DC-conductance
contribution associated with combined EV and Majorana tunneling processes to survive moderate
asymmetry, provided coherence is maintained. A quantitative treatment of the fully
asymmetric X-shaped interferometer is beyond the scope of the present work but could
benefit from the technical advances discussed in Ref.~\cite{giuliano2026}.

Finally, let us discuss the length and temperature scales relevant to an experimental realization of the proposed device. Assuming representative parameters for $v$ and $\Delta$, see Sec.~\ref{sec2a}, the localization length of the Majorana edge states is $\ell_c=\hbar v/\Delta\simeq 16$~nm. SC strips with widths of order $W\approx 100$~nm or larger  therefore strongly suppress any direct overlap between modes on opposite edges. In contrast, the Josephson weak links can locally bring the modes sufficiently close to generate appreciable Majorana tunneling amplitudes. The required dimensions depend on the junction transparency and the microscopic decay lengths; typical widths are of the order $10~$nm.
Thermal averaging introduces the length scale
$\ell_T=\frac{\hbar v}{\pi k_BT}\simeq 0.24~\mu\mathrm{m} \frac{100~\mathrm{mK}}{T}$. 
Therefore, temperatures of $50$ to $100$~mK correspond to thermal lengths of approximately 
$0.5$ to $0.25~\mu$m, suggesting compact devices with relevant propagation distances in 
the submicron range. For interfering paths with unequal propagation times, 
avoiding substantial thermal averaging requires their time difference to be small 
compared with $\hbar/(\pi k_B T)$; for equal velocities, this corresponds to a path-length mismatch $|\delta L|\lesssim\ell_T$. This estimate is a design guideline rather than a 
quantitative result for a fully asymmetric X geometry. 
In addition, coherent propagation requires the device dimensions to remain below the inelastic coherence length which is distinct from $\ell_T$.
Furthermore, the island must realize the Josephson-dominated regime $E_J\gg E_C$ assumed here while retaining a measurable quantum-phase-slip amplitude. Increasing the ratio $E_J/E_C$ suppresses phase slips exponentially, i.e., this ratio cannot be made arbitrarily large without reducing the signal. In addition, the thermal and bias energies must remain below the excitation gap underlying the effective edge-state description. For the above SC gap, $\Delta/k_B\simeq 4.8$~K, temperatures in the proposed range comfortably satisfy $k_B T \ll\Delta$. 
These estimates identify a plausible range of device dimensions and operating temperatures. The achievable conductance magnitude will additionally depend on the junction tunneling amplitudes, phase-slip amplitude and decoherence, requiring device-specific characterization.

To conclude, multi-terminal interferometer geometries harboring chiral Majorana modes can provide conceptually simple, and hopefully soon realizable, setups which allow experimentalists to probe the  non-Abelian statistics of Ising anyons in terms of easily accessible linear-response DC transport observables.

\section*{Data Availability}

The data underlying all figures in this paper are available on Zenodo \cite{Zenodo}.

\section*{Acknowledgments}

We thank Anton Akhmerov and Carlo Beenakker for discussions.
We acknowledge funding by the Deutsche Forschungsgemeinschaft (DFG, German Research Foundation) under Projektnummer 277101999 - TRR 183 (project C01), under Projektnummer EG 96/15-1, and under Germany's Excellence Strategy - Cluster of Excellence Matter and Light for Quantum Computing (ML4Q) EXC 2004/2 - 390534769.

 
\bibliographystyle{jsty3-author}
\bibliography{refs}
 
\end{document}